\documentclass[a4paper, 10pt, conference]{IEEEtran}

\usepackage[dvipsnames]{xcolor}
\usepackage{url,hyperref}
\usepackage{mathtools, cuted}
\usepackage[noadjust, nobreak]{cite}
\usepackage{tabularx}

\title{\LARGE \bf
A Gateway to Astronomical Image Processing: Vera C. Rubin Observatory LSST Science Pipelines on AWS
}

\author{\IEEEauthorblockN{
    Dino Bektesevic$^{1,a}$, 
    Hsin-Fang, Chiang$^{2,3}$,
    Kian-Tat Lim$^{2,4}$,
    Todd L. Miller$^{6}$, 
    Greg Thain$^{6}$, \\
    Tim Jenness$^{2,3}$, 
    James Bosch$^{2,5}$,
    Andrei Salnikov$^4$,
    Andrew Connolly$^{1,b}$}\\
    \footnotesize
    $1.$ Department of Astronomy, University of Washington; 
    $2.$ Vera C. Rubin Observatory; 
    $3.$ AURA;
    $4.$ SLAC National Accelerator Laboratory\\
    $5.$ Department of Astrophysical Sciences, Princeton University;
    $6.$ Center for High-Throughput Computing at the University of Wisconsin-Madison \\
    \footnotesize
    Email: $a.$ dinob@uw.edu, $b.$ ajc@astro.washington..edu
    }

\IEEEoverridecommandlockouts

\begin{document}

\maketitle

%%%%%%%%%%%%%%%%%%%%%%%%%%%%%%%%%%%%%%%%%%%%%%%%%%%%%%%%%%%%%%%%%%%%%%%%%%%%%%
\begin{abstract}
The Legacy Survey of Space and Time \cite{Ivezic2019}, operated by the Vera C. Rubin Observatory, is a 10-year astronomical survey due to start operations in 2022 that will image half the sky every three nights. LSST will produce  $\sim$20TB of raw data per night which will be calibrated and analyzed in almost real time.  Given the volume of LSST data, the traditional subset-download-process paradigm of data reprocessing faces significant challenges. We describe here, the first steps towards a gateway for astronomical science that would enable astronomers to analyze images and catalogs at scale. In this first step we focus on executing the Rubin LSST Science Pipelines, a collection of image and catalog processing algorithms, on Amazon Web Services (AWS). We describe our initial impressions on the performance, scalability and cost of deploying such a system in the cloud.
\end{abstract}

%%%%%%%%%%%%%%%%%%%%%%%%%%%%%%%%%%%%%%%%%%%%%%%%%%%%%%%%%%%%%%%%%%%%%%%%%%%%%%
\section{Introduction}
The currently pervasive model of sub-selecting, transferring to local compute resources and then reprocessing data has been successful for users processing past astronomical sky-survey data because technological developments and pricing made acquiring sufficient local compute resource affordable. With a new generation of sky surveys such as those operated by the Rubin Observatory delivering an order of magnitude more data this subset-download-process paradigm is no longer viable for users of these datasets. We describe a different approach, utilizing the elastic capabilities of cloud compute resources to enable users to scale their analyses to 100TB+ data sets.

\begin{figure}[htb]
\centering
\includegraphics[angle=-90,width=0.8\columnwidth]{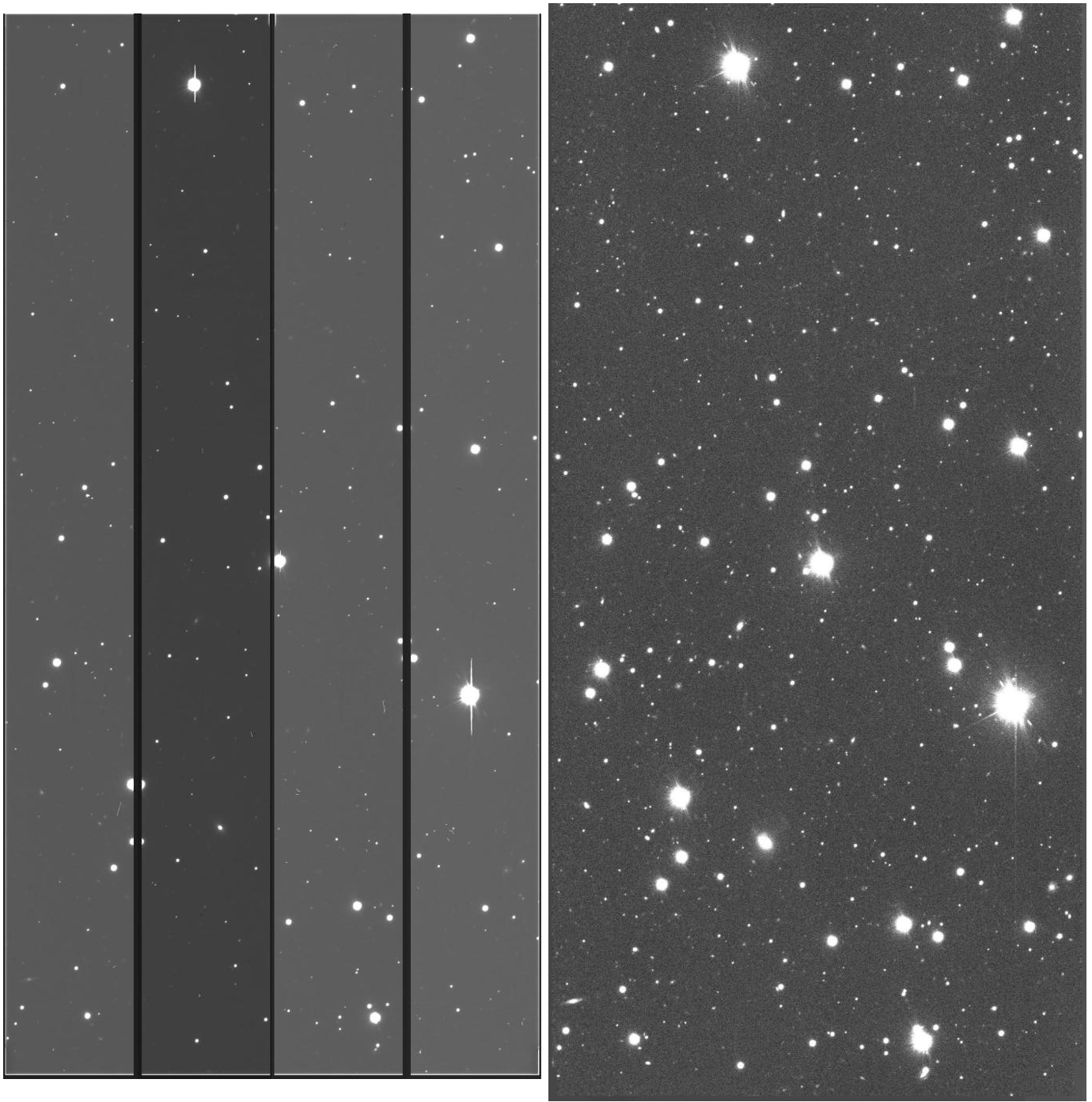}
\caption{A raw, uncalibrated image from the Hyper Suprime Cam (HSC) \cite{HSC} (top) and the resulting calibrated exposure after processing with the Rubin LSST Science Pipelines (bottom). The images are not the same size because calibration adjusts for the optic distortion of the instrument.}
\label{fig:raw2calexp}
\end{figure}

Our goal is to provide astronomers with an interface and tools that allow them to reprocess and reanalyze an entire night's worth of Rubin data in hours, and to do so at a reasonable cost. As a first step towards such a gateway we focus on image data reduction pipelines. These pipelines consist of a series of steps that remove any instrumental signature from the data, detect sources above a specified threshold, potentially cross reference these detections to previously known sources, and measure their properties (or features). A typical input and output of such a pipeline is shown on Figure \ref{fig:raw2calexp} where a raw image has been processed to remove instrumental effects. Exposing the Rubin LSST Science Pipelines functionality through a common interface would enable the astronomy community
to define and execute custom analysis pipelines based on state-of-the-art astronomical data processing algorithms. The need for scalability and the ability to share the resulting data across a range of communities naturally leads to a cloud compute model whether commercial or academic.

%%%%%%%%%%%%%%%%%%%%%%%%%%%%%%%%%%%%%%%%%%%%%%%%%%%%%%%%%%%%%%%%%%%%%%%%%%%%%%
\section{Technology Stack}
\label{sec:techstack}
While cloud technology has been embraced quickly by some scientific communities (see \cite{NatLang} or \cite{HEPCloud}) its adoption by the astronomical community has been slow. Legacy code is typically designed with the assumptions that (a) data exists locally or there exists a globally accessible file system, (b) the computation is some form of batch processing, and (c) the system is in general not state agnostic. While it is possible to create similar systems in the cloud, modern cloud approaches scale much better with shared-nothing filesystems \cite{sharednothing}, containerization \cite{containerization} and near-data processing architectures. 

Here we will firstly provide a brief overview of Rubin LSST Science Pipelines and how they are used to process data and then we will briefly describe the modifications required to run the system in the cloud. 

The Rubin LSST Science Pipelines \cite{sciencepipelines} represent the state-of-the-art in astronomical data reduction. They consist of configurable Tasks that can be chained into a pipeline. Such a pipeline is described by a directed acyclic graph (DAG) called a Quantum Graph \cite{dmtn055}. A Quantum Graph consists of Quanta, where each Quantum is a Task applied to an individual dataset. The Rubin LSST Science Pipelines enable processing of astronomical data from a single exposure to overarching tasks such as joint calibration that constrains astrometric and photometric measurements across multiple exposures. 

Tasks themselves are agnostic to file formats and locations of the data. The input-output (IO) and provenance is tracked through a Middleware component called the Data Butler \cite{databutler}. The main purpose of the Data Butler is to isolate the end user from file organization, filetypes and related file access mechanisms by exposing datasets as, mostly, Python objects. Datasets are referred to by their unique IDs, or a set of identifying references. The Data Butler uses a registry to resolve the dataset references and resolves the location, file format and the Python object/type of the files stored in a datastore.

\href{https://research.cs.wisc.edu/htcondor}{HTCondor} \cite{Thain:2005:HTCondor} provides a powerful batch system for high throughput computing (HTC). Directed Acyclic Graph Manager (DAGMan) is HTCondor's metascheduler capable of managing workflows at a higher level than the underlying HTCondor Scheduler. \href{https://pegasus.isi.edu/}{Pegasus} \cite{Deelman:2015:Pegasus} is a workflow management system built on top of HTCondor. It provides command line and API interfaces that allow writing abstract workflows (DAXs) independent of the underlying computing infrastructure. Pegasus then translates the abstract workflow to an executable workflow interpretable by DAGMan. The Rubin LSST Science Pipelines Quantum Graph is written in the form of a DAX, that is submitted to Pegasus, which translates it into an executable workflow and submits it to HTCondor. 

\begin{figure}[htb]
\centering
\includegraphics[width=0.9\columnwidth]{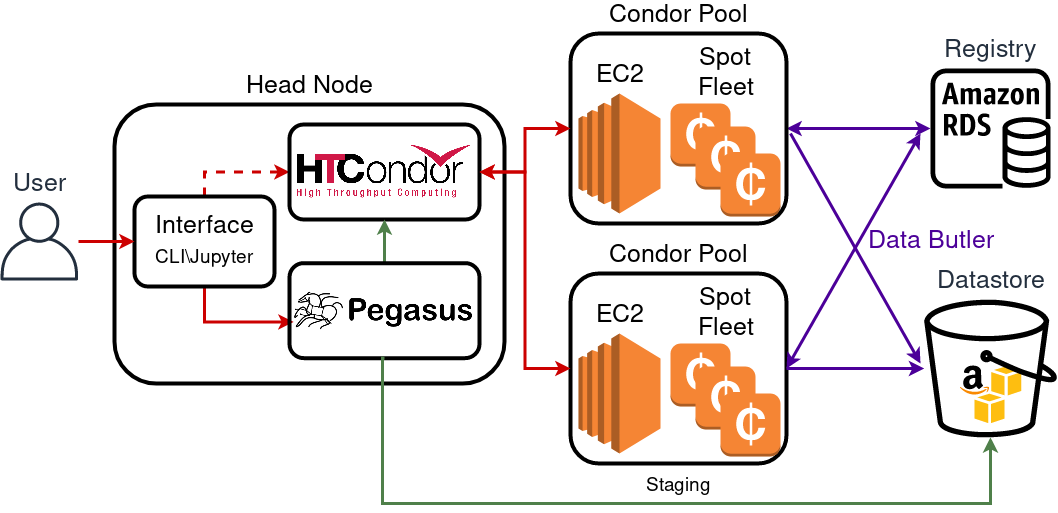}
\caption{A user or users, leftmost, gains access to the head node which contains the Rubin LSST Science Pipelines, Pegasus and HTCondor. The user creates and submits a Quantum Graph to Pegasus. Pegasus clusters the jobs and begins submitting jobs to HTCondor. Users procure compute resources, workers, through HTCondor Annex. Compute resources are logically separated into units called Condor Pools, each of which is dynamically scalable and can consist of a mix of EC2 instances procured on-demand or via Spot Fleet requests. As Quanta are executed on the workers the persisted data is written to the Datastore, an S3 Bucket, and to the Registry, an RDS PostgreSQL database. Persisted data remains accessible from the head node.}
\label{fig:techstack}
\end{figure}

To facilitate running Rubin LSST Science Pipelines in the cloud we adopted an lift-and-shift strategy that is common in porting existing applications to the cloud. Initially an exact copy of Rubin software was executed on cloud services configured to mimic the on-premise environment. Having verified functionality of the software we implemented an \href{https://aws.amazon.com/s3/}{Simple Storage Service (S3)} back-end for the Datastore and a PostgreSQL backed Registry managed through the \href{https://aws.amazon.com/rds/}{Relational Database Service (RDS)} service. Because the Data Butler abstracts IO for the Middleware, the changes remove the need to have a shared local drive across the nodes. HTCondor Annex allows HTCondor deployment on cloud resources via acquisition of cloud compute resources external to an existing HTCondor pool by acquiring Elastic Compute 2 (EC2) instances. The schematic of the system is shown on Figure \ref{fig:techstack}.

Scaling is achieved by scheduling as many parallel jobs of the same type as resources allow. When the compute cluster is under-subscribed due to a low number of jobs to schedule simultaneously, the resource manager will deallocate the idle resources for cost optimization. New resources can also be allocated and dynamically added to the cluster when resource requirements of the jobs change. The system, and work required to create it, are described in \cite{dmtn114} and \cite{dmtn135}. In the following chapters we focus on analysing and identifying bottlenecks and mitigation strategies that lead to a decrease in cost and an increase in performance.

%%%%%%%%%%%%%%%%%%%%%%%%%%%%%%%%%%%%%%%%%%%%%%%%%%%%%%%%%%%%%%%%%%%%%%%%%%
\section{Example Workflow}
\label{sec:exampleWorkflow}
\begin{figure}[htb]
\centering
\includegraphics[width=0.9\columnwidth]{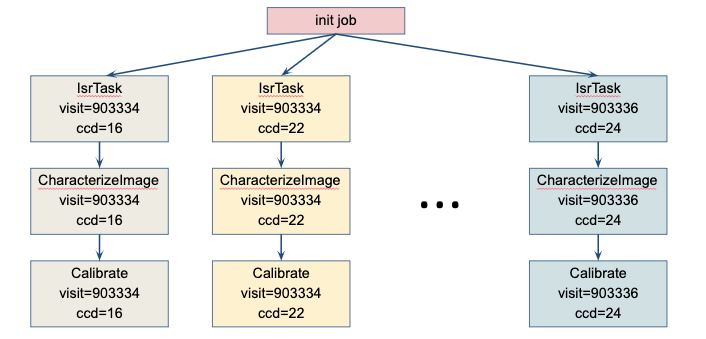}
\caption{An example workflow for performance and scalability tests. Processing consists of instrument signature removal, characterization and calibration. These are ordered jobs for each image but processing a set of images is an embarrassingly parallel task.}
\label{fig:demo-workflow}
\end{figure}

The example workflow is based on the HSC Release Candidate (RC) dataset tract 9516  \cite{dmtr31}. This dataset is reprocessed using Rubin compute resources every two weeks in order to characterize the scientific validity and performance of the algorithms. The dataset contains 6787 images. The total number of Quanta, i.e.\ jobs, is 20361. There is one initialization task that prepares the Registry and there are three different Tasks that run on the images: Instrument Signature Removal (ISR), Characterization, and Calibration. A broad overview of the Tasks is provided in \cite{sciencepipelines} with more details provided in \cite{Bosch_2017}. Processing results in an instrument corrected image, a model of the background and the point spread function, and a catalog of detected sources on which astrometric and photometric measurements were performed. The workflow is shown on Figure \ref{fig:demo-workflow}.
The total size of the input data is 0.2TB and the size of output data is approximately 2.7TB. In number of exposures processed this represents $\sim$11\% of the number of observations per night from Rubin Observatory.

%%%%%%%%%%%%%%%%%%%%%%%%%%%%%%%%%%%%%%%%%%%%%%%%%%%%%%%%%%%%%%%%%%%%%%%%%%
\section{Performance, scalability and cost}
The example workflow was evaluated for multiple different instance types and instance sizes. The tests were performed using on-demand and spot instance types. The wall time is the cumulative value of execution times of all Quanta in the Quantum Graph, including the scheduling and execution overhead. Each workflow includes the initialization task which, on average, adds an overhead of 25 minutes.

Initial testing was performed with 25, 50 and 100 \texttt{c5.xlarge} instances. Each \texttt{c5.xlarge} instance has 4 vCPUs. Each vCPU is a hardware hyperthread on a 3.0 GHz Intel Xeon Platinum 8000-series processor, 8 GB of memory, up to 10 Gbps connection bandwidth and mounts an General Purpose (\texttt{gp2}) Elastic Block Storage (EBS) SSD volume. The two major bottlenecks in the processing were an EBS drive IO bottleneck and the staging of the scripts and configuration files.

A typical \texttt{gp2} EBS drive throughput is between 128 and 250 MBps, i.e. on par with modern HDDs, depending on volume size and instance type. This is usually parameterized through IOPS (IO operations per second) with larger volumes having a higher base IOPS performance, e.g. a 100 GB drive's base performance is 300 IOPS. A volume can burst up to 3000 IOPS with the length of the burst governed by the amount of Burst Credits associated with the drive. Burst Credits recharge when IOPS usage is less than baseline performance. The Data Butler can not always instantiate objects from memory and will instead download files to the local drive, which can exhaust the Burst Credits on individual workers. 

\begin{figure}[htb]
\centering
\includegraphics[width=0.9\columnwidth]{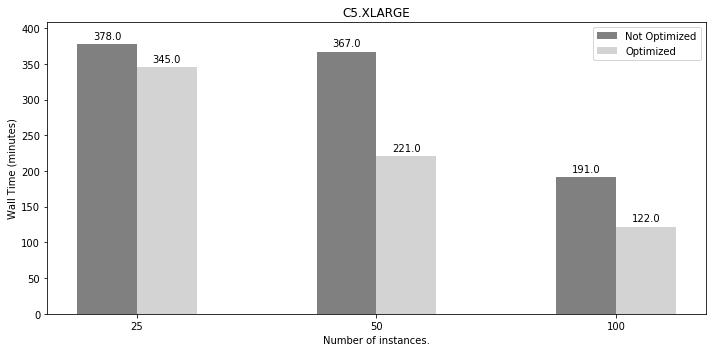}
\caption{Workflow execution wall time on \texttt{c5.xlarge} instances before and after optimization.}
\label{fig:c5WallT}
\end{figure}

Staging is the process of setting up the required environment including the transfer of job scripts, their configuration files and other required files. We stage the processing configuration and output logs, using Pegasus, to and from the workers respectively. The cumulative size of the processing configuration files can be up to 5 GB and 6 GB for output processing logs, straining the IO performance on the head node as many of the configuration and output logs transfers occur in lock-step. During testing, Burst Credits often masked or delayed the onset of IO bottle-necking, making it hard to identify the issue. For example, timings for 25 and 50 workers were the same (see Figure-\ref{fig:c5WallT} unoptimized) because the Burst Credits were already exhausted before the first run, so both workflows were limited by the drive throughput, while the third one, launched the following day, executed against full Burst Credits.

Selecting more appropriately scaled EBS drives for the head and worker nodes, in tandem with moving the staging site from the head node and into an S3 Bucket are directly responsible for the majority of the measured improvement. Further improvements were achieved by clustering multiple single jobs into a single larger job and by relegating scheduling to DAGMan. This was because, Pegasus would, by default, fork a new \texttt{condor\_submit} process for every job. If the workflow was, for example, executed on 50 \texttt{c5.2xlarge} workers it would launch 400 jobs, which fork 400 processes and establish 400 head-worker staging transfer connections, simultaneously.

Better indexing schemes for the Registry, materialized views, allocating more cache per connection, increasing the maximum number of allowed connections and simultaneous locks resulted in further a decrease in walltime and cost of the workflows.

%%%%%%%%%%%%%%%%%%%%%%%%%%%%%%%%%%%%%%%%%%%%%%%%%%%%%%%%%%%%%%%%%%%%%%%%%%
\subsection{Resource Optimization}
\begin{figure}[htb]
\centering
\includegraphics[width=0.9\columnwidth]{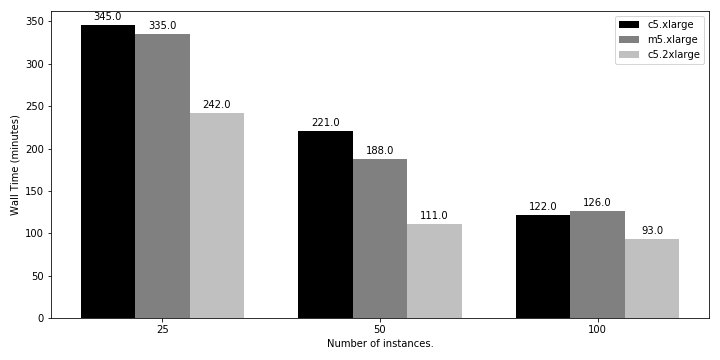}
\caption{EC2 wall times for the example workflow.
}
\label{fig:AllWallT}
\end{figure}

\begin{figure}[htb]
\centering
\includegraphics[width=0.9\columnwidth]{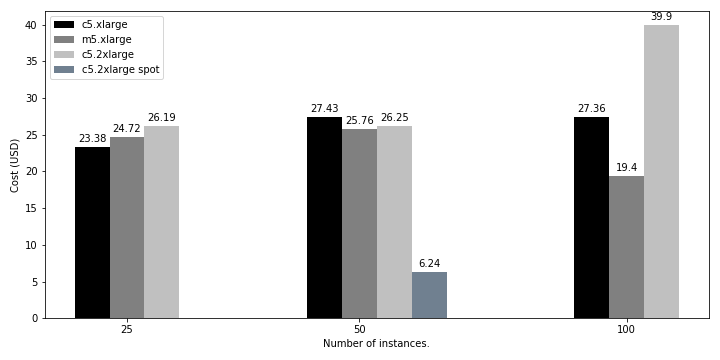}
\caption{EC2 cost for the example workflow.}
\label{fig:AllCost}
\end{figure}

Choice of instance type, number of instances and workflow determines the cost (either in wall time or USD) for the execution of the Rubin LSST Science Pipelines. In Figure-\ref{fig:AllWallT} we show the performance scaling tests after workflow optimization. The example workflow shows scaling that is approximately linear with the number of simultaneously allocated jobs, which is dependent on the number of vCPUs, memory and job resource requirements. The scaling factor of 1.8 is only slightly less than what is expected from an idealized case. The 50 and 100 workers tests for \texttt{c5.2xlarge} family do not show linear scaling. The total duration was 1h18m when executing with 100 workers, meaning the intialization task became a significant portion of the total wall time. Additionally, the example workflow does not have sufficient parallelizable jobs to saturate the allocated resources, i.e.\ the compute cluster was undersubscribed. Because resource allocation is based on vCPU (which are hyperthreads for the selected instance families and sizes) it is still possible for the physical core to be occupied and therefore the worker will not deallocate automatically but its full resources will remain unused. To verify we executed an additional test using 200 \texttt{m5.large} workers, counting 400 vCPUs with 2 GB per vCPU of memory. The test finished in approximately the same wall time, 1h58m, as its vCPU number counterpart, the 50 \texttt{c5.2xlarge} workers. The cost for the workflows was 28.21 USD and 26.72 USD respectively, which confirmed that the scaling is mainly dependent on the number of jobs that are able to run in parallel.
The cost of EC2 instances is shown in Figure-\ref{fig:AllCost}. We find that cost is approximately constant, within 7\%, for tests using 25 and 50 workers. This is to be expected as the cost of more powerful instances is offset by a reduction in wall time. Therefore, if the cluster is oversubscribed, increasing the number of workers decreases both the wall time and the cost of the workflow (if the instances suit the workflow resource requirements). The workflow executed on 100 \texttt{c5.2xlarge} workers costs significantly more because the cluster was under-subscribed but, as explained previously, it was not possible to deallocate workers. The workflow executed on 100 \texttt{m5.xlarge} instances costs significantly less than its 25 and 50 worker counterparts because the reduction in wall time scaled as 1.8 meaning the instances were allocated for shorter periods of time. Significant savings are achieved by executing on Spot instances, with our example workflow, using 50 \texttt{c5.2xlarge} workers, reduced by a factor $>4$ (see Figure-\ref{fig:AllCost}). The total workflow cost is also offset by the constant cost of the RDS instance, the head node and amount of data stored in S3.
In general the scaling performance is not limited by IO or connection bandwidth on the workers mainly due to the distributive properties of S3. The upper limit is generally set by the size of the instance that hosts the Registry. Since all jobs need to contact the registry, there is an  upper limit to the number of workers that can execute simultaneously set by the maximum number of simultaneous connections and maximum allowed number of locks per connection, that are in turn dependant on the available cache size of the RDS instance. The largest RDS instance available is the \texttt{db.m5.24xlarge} with 96 vCPUs, 384 GB of memory and 14000 Mbps of write speed to the attached EBS drives. The per-hour cost of such an instance is 8.45 USD/h of use which is substantial in an on-demand provisioning scheme. Were this functionality offered as a service, however, cost can be significantly amortized by procuring reserved instances. For reserved instances a 40\% discount is possible by reserving an instance for a year and 60\% for three years.

\section{Cost projections}
The dataset used in this analysis represents approximately 11\% of the number of  nightly observations of Rubin or, given the difference in the size of the cameras, about 2\% of Rubin's nightly data volume. In \cite{dmtn137} we performed a cost estimate of a full Data Release Production (DRP) workflow, a workflow that in addition to the tasks described in \ref{sec:exampleWorkflow} also includes combining images (coaddition) and source measurements on the combined images.

We base our on-premise resources estimates on \cite{dmtn135} where we multiply the per node cost with the walltime for a workflow executed using on-premise resources. The Verification Cluster at the National Center for Supercomputing Applications (NCSA) is used as a prototypical Rubin compute center. The Verification Cluster comprises 48 Dell C6320 nodes. Each node has 24 physical cores and 128 GB RAM, which due to hyperthreading results in 48 virtual cores per node. We estimate the cost of the on-premise compute resources to be 0.408 USD per node-hour without including labor. We note that until the location of the Rubin Data Facility has been decided and concrete commitments are made this costing estimate will remain uncertain. 

To provide a bound on the expected requirements and costs, in the following analysis we evaluate the DRP workflow using two different data abstraction layers known as the Generation 2 and 3 Butlers. Generation 2 Butler is the current and more mature implementation, and Generation 3 Butler, on which our work is based, is the less optimized next generation Data Butler. Processing took 17.8 node hours \cite{dmtn160} and 88 node hours on the Verification Cluster for Data Butler Generation 2 and 3 respectively. This corresponds to run time costs of 7.3 USD and 35.9 USD. Given the large difference in the run times for the different generations, to estimate the total cost of processing Rubin data with the DRP pipelines, we use the {\it relative} performance of running these workflows in the cloud and on-premises and use that to scale the costs given in \cite{dmtn137}.

Rubin DRP estimates that approximately 2PB of raw science data will be produced annually. Based on an input size of 122 GB and a cost of 95 USD for the DRP workflow \cite{dmtn137}, assuming that the scaling remains linear, Rubin processing would cost approximately 0.7 million USD for the first year of DRP operations if we adopted no cloud optimizations. This compares to 0.9 million USD for the current on-premises implementation. If the optimizations we achieved in this paper can be extended across the entire DRP workflow, not just the example workflow we have evaulated, we estimate that Rubin yearly DRP processing costs would be between 600,000 to 900,000 USD. Note that this includes the compute, database access, and storage necessary to run the pipelines. It does not include the long-term archival costs of storing or querying the raw and processed images or the resulting catalogs. From these numbers, it is clear that optimization can yield substantial cost savings (between a factor of 1.7 and 2.5). If the optimizations shown in this paper can be extended across the full DRP workflow we believe that the compute resources required for processing Rubin data could be cost neutral on the cloud.

%%%%%%%%%%%%%%%%%%%%%%%%%%%%%%%%%%%%%%%%%%%%%%%%%%%%%%%%%%%%%%%%%%%%%%%%%%%%%%
\section{Conclusions}
We show that a lift-and-shift strategy is an appropriate and relatively easy way to adapt Rubin software to run in the cloud. While an unoptimized lift-and-shift approach can scale to multi-petabyte datasets, such as those that will be generated by  Rubin Observatory, it is not a cost effective solution. While the current work does not address storage or networking costs, we identify several compute performance bottlenecks that drive processing costs and propose solutions to mitigate them. The most significant compute performance gains are achieved by careful management of workflow execution IO. Trimming significant amounts of log transfers, staging in an S3 bucket, and clustering reduced the example workflow walltime from 210--255 minutes to 128--140 minutes when executed on 50 \texttt{c5.xlarge} instances. The large dispersion in the unoptimized lift-and-shift example workflow walltimes is due to the nature of some of the bottlenecks. 

We investigate the impact of EC2 instance size and type on performance and cost. We show that compute resources must be tailored to the requirements of individual Tasks within a complex workflow. For the example workflow, for small numbers of instances, walltime is a function of instance type and instance size and walltime scales linearly with the number of jobs that can run in parallel. Above 50 instances, this scaling no longer holds because the example dataset is not large enough to saturate the compute resources of the workers. Costs remain flat relative to instance type and number of instances for $<50$ instances but for larger instance counts the under utilization of the compute resources increases the total cost. While the cost estimates for workflows run on inappropriately selected compute resources might not differ significantly, due to different pricing of instances, the total wall time of such a workflow will. Selecting a cost and performance optimal instance type and size requires careful benchmarking of each individual component of a workflow. For the case of the Rubin pipelines, adding more fine grained control that would allow associating, launching, or at least targeting Quanta to specific resources would enable better resource allocation, improve total cost, and reduce run times.

Lastly, while the current results might not match the cost estimates for purchasing large scale on-premise compute resources for a long term 10-year sky survey such as the LSST, we show that it is possible to approach those estimates within 30\% to 40\%. We believe that for such a difference in relative pricing between on-premise and on-cloud resources, the added benefits of elasticity could outweigh the absolute cost of the workflow. Based on the work presented, we believe it is possible to achieve this relative price difference even when not including different provisioning schemes other than on-demand and spot. For longer lasting projects it would also be possible to reserve the EC2 and RDS resources required with 1 to 3 year reserved instance contracts. The pricing of EC2 and RDS instances is then further reduced by up to 72\% and 65\% respectively. With further work on the optimization of the workflow and system performance we believe, for the compute resources, cost parity with on-premises solutions can be achieved. 

We note that the elasticity provided by moving into the cloud could allow the full Rubin data set to be reprocessed in under a week rather than the 6 months expected from an on-premises solution. Supplementing on-premise resources with on-demand or spot compute resources in the cloud could improve the speed with which research is conducted yielding a total net saving in delivering science results. This would transform how quickly science discoveries could be made using Rubin data. \\

\section{Acknowledgements}
\noindent We would like to thank Rubin Observatory and the AWS PoC group, which were instrumental in writing and testing much of the code to facilitate AWS processing. We acknowledge support from NSF Award OAC-1739419.
%%%%%%%%%%%%%%%%%%%%%%%%%%%%%%%%%%%%%%%%%%%%%%%%%%%%%%%%%%%%%%%%%%%%%%%%%%%%%%
\bibliographystyle{ieeetr}
\bibliography{main}

\begin{thebibliography}{10}

\bibitem{Ivezic2019}
{\v{Z}}.~Ivezi\'{c}, S.~M. Kahn, J.~A. Tyson, and et~al., ``{LSST}: {F}rom
  {S}cience {D}rivers to {R}eference {D}esign and {A}nticipated {D}ata
  {P}roducts,'' {\em The Astrophysical Journal}, vol.~873, p.~111, Mar 2019.

\bibitem{HSC}
H.~{Aihara}, N.~{Arimoto}, R.~{Armstrong}, and et~al., ``{The Hyper Suprime-Cam
  SSP Survey: Overview and survey design},'' {\em Publications of the
  Astronomical Society of Japan}, vol.~70, p.~S4, Jan. 2018.

\bibitem{NatLang}
B.~{Posey}, C.~{Gropp}, B.~{Wilson}, and et~al., ``{Addressing the Challenges
  of Executing a Massive Computational Cluster in the Cloud},'' in {\em 2018
  18th IEEE/ACM International Symposium on Cluster, Cloud and Grid Computing
  (CCGRID)}, pp.~253--262, 2018.

\bibitem{HEPCloud}
B.~{Holzman}, L.~A.~T. {Bauerdick}, B.~{Bockelman}, and et~al., ``{HEPCloud, a
  New Paradigm for HEP Facilities: CMS Amazon Web Services Investigation},''
  {\em arXiv e-prints}, p.~arXiv:1710.00100, Sept. 2017.

\bibitem{sharednothing}
M.~Stonebraker, ``The case for shared nothing,'' {\em {IEEE} Database Eng.
  Bull.}, vol.~9, no.~1, pp.~4--9, 1986.

\bibitem{containerization}
D.~{Bernstein}, ``Containers and cloud: From {LXC} to {D}ocker to
  {K}ubernetes,'' {\em IEEE Cloud Computing}, vol.~1, no.~3, pp.~81--84, 2014.

\bibitem{sciencepipelines}
J.~Bosch, Y.~AlSayyad, R.~Armstrong, and et~al., ``{A}n {O}verview of the
  {LSST} {I}mage {P}rocessing {P}ipelines,'' 2018.

\bibitem{dmtn055}
G.~Dubois-Felsmann, ``{S}uper{T}ask {A}rchitecture and {D}esign.''
  https://dmtn-055.lsst.io, 2017.

\bibitem{databutler}
T.~Jenness, J.~F. Bosch, P.~Schellart, and et~al., ``{A}bstracting the storage
  and retrieval of image data at the {LSST},'' 2018.

\bibitem{Thain:2005:HTCondor}
D.~Thain, T.~Tannenbaum, and M.~Livny, ``{D}istributed computing in practice:
  the {C}ondor experience,'' {\em Concurrency and Computation: Practice and
  Experience}, vol.~17, no.~2‐4, pp.~323--356, 2005.

\bibitem{Deelman:2015:Pegasus}
E.~Deelman, K.~Vahi, G.~Juve, and et~al., ``{P}egasus: a {W}orkflow
  {M}anagement {S}ystem for {S}cience {A}utomation,'' {\em {F}uture
  {G}eneration {C}omputer {S}ystems}, vol.~46, pp.~17--35, 2015.

\bibitem{dmtn114}
K.-T. Lim, L.~Guy, , and H.-F. Chiang, ``{LSST} + {AWS} {P}roof of {C}oncept.''
  \url{https://dmtn-114.lsst.io}, 2019.

\bibitem{dmtn135}
M.~Butler, K.~T. Lim, and W.~O’Mullane, ``{DM} sizing model and purchase plan
  for the remainder of construction.'' \url{https://dmtn-135.lsst.io}, 2020.

\bibitem{dmtr31}
H.-F. Chiang and M.~W. Johnson, ``{A}s-is {HSC} {R}eprocessing.''
  https://docushare.lsst.org/docushare/dsweb/Get/DMTR-31, 2017.

\bibitem{Bosch_2017}
J.~Bosch, R.~Armstrong, S.~Bickerton, and et~al., ``{T}he {H}yper
  {S}uprime-{C}am software pipeline,'' {\em Publications of the Astronomical
  Society of Japan}, vol.~70, Oct 2017.

\bibitem{dmtn137}
H.-F. Chiang, D.~Bektesevic, and the AWS-PoC~team, ``{AWS} {P}roof of {C}oncept
  {P}roject {R}eport.'' \url{https://dmtn-137.lsst.io}, 2020.

\bibitem{dmtn160}
H.-F. Chiang and S.~Thrush, ``{S}18 {HSC} {PDR}1 reprocessing.''
  https://dmtn-160.lsst.io, 2020.

\end{thebibliography}
%\section{Acknowledgements}
%\noindent We would like to thank Rubin Observatory and the AWS PoC group which were instrumental in testing, accepting and committing to maintaining much of the code written that was written to facilitate AWS processing. We acknowledge support from NSF Award OAC-1739419.
\clearpage
\end{document}